\pdfoutput=1
\newif\ifarxiv
\arxivtrue

\ifarxiv
  \documentclass[manuscript,screen,nonacm]{acmart}
\else
  \documentclass[manuscript,screen,review,anonymous]{acmart}
\fi
\newcommand{\blinded}[2]{\ifarxiv#2\else#1\fi}

\usepackage{booktabs}
\usepackage{longtable}
\usepackage{amsmath}

\graphicspath{{figures/}}

\acmJournal{TOCHI}
\acmVolume{0}
\acmNumber{0}
\acmArticle{0}
\acmMonth{0}
\acmYear{2026}
\copyrightyear{2026}
\setcopyright{none}

\newcommand{\NAnnotated}{527}
\newcommand{\NDropped}{21}
\newcommand{\PctDropped}{4.0}
\newcommand{\NAnalytic}{506}
\newcommand{\NUseCaseCoded}{474}
\newcommand{\NValCoded}{470}
\newcommand{\PctUS}{97.6}
\newcommand{\PctHigherEd}{93.7}

\newcommand{\PctMen}{53.6}
\newcommand{\PctWomen}{42.1}

\newcommand{\PctStudentRA}{40.5}

\newcommand{\PctFaculty}{16.6}
\newcommand{\PctLifeSci}{28.7}
\newcommand{\PctEngineering}{20.6}
\newcommand{\PctSocialSci}{10.7}
\newcommand{\PctPhysSci}{10.3}

\newcommand{\MedianYears}{6}
\newcommand{\IqrYearsLo}{4}
\newcommand{\IqrYearsHi}{10.5}
\newcommand{\MaxYears}{50}
\newcommand{\PctWeeklyPlus}{84}
\newcommand{\PctDaily}{50.2}
\newcommand{\PctWeekly}{34.2}
\newcommand{\PctPython}{70.4}
\newcommand{\PctR}{56.3}
\newcommand{\PctMatlab}{30.8}
\newcommand{\PctBash}{20.6}
\newcommand{\PctCpp}{15.0}
\newcommand{\PctChatGPT}{61.7}
\newcommand{\PctCopilot}{11.7}
\newcommand{\PctCustomTool}{5.7}
\newcommand{\PctGemini}{5.5}
\newcommand{\PctClaude}{4.3}
\newcommand{\PctVersionControl}{50.7}
\newcommand{\PctCodeReview}{33.6}
\newcommand{\PctUnitTestPractice}{28.7}
\newcommand{\PctSystemTestPractice}{18.6}
\newcommand{\PctRegressionTestPractice}{17.6}
\newcommand{\NUseCaseTypes}{13}
\newcommand{\NDataHandling}{121}
\newcommand{\PctDataHandling}{23.9}
\newcommand{\NVisualization}{100}
\newcommand{\PctVisualization}{19.8}
\newcommand{\NDebugging}{88}
\newcommand{\PctDebugging}{17.4}
\newcommand{\NMathSci}{59}
\newcommand{\PctMathSci}{11.7}
\newcommand{\NAnalysisUC}{55}
\newcommand{\PctAnalysisUC}{10.9}
\newcommand{\NSystemsHw}{49}
\newcommand{\PctSystemsHw}{9.7}
\newcommand{\NTranslation}{24}
\newcommand{\PctTranslation}{4.7}
\newcommand{\NCodeComp}{23}
\newcommand{\PctCodeComp}{4.5}
\newcommand{\NOptimization}{21}
\newcommand{\PctOptimization}{4.2}
\newcommand{\NUserInterface}{14}
\newcommand{\PctUserInterface}{2.8}
\newcommand{\NRefactor}{11}
\newcommand{\PctRefactor}{2.2}
\newcommand{\NValidationAct}{9}
\newcommand{\PctValidationAct}{1.8}
\newcommand{\NDocumentation}{7}
\newcommand{\PctDocumentation}{1.4}
\newcommand{\PctTopFiveUseCaseShare}{76}

\newcommand{\PctSingleUseCase}{75.7}
\newcommand{\MeanUseCaseCodes}{1.15}
\newcommand{\NOneStrategy}{249}
\newcommand{\PctOneStrategy}{49.2}
\newcommand{\NTwoStrategies}{176}
\newcommand{\PctTwoStrategies}{34.8}
\newcommand{\NThreePlusStrategies}{45}
\newcommand{\PctThreePlusStrategies}{8.9}
\newcommand{\MeanValCodes}{1.47}
\newcommand{\NRunCode}{267}
\newcommand{\PctRunCode}{52.8}
\newcommand{\NInspectOutput}{114}
\newcommand{\PctInspectOutput}{22.5}
\newcommand{\NReadCode}{98}
\newcommand{\PctReadCode}{19.4}
\newcommand{\NInspectViz}{82}
\newcommand{\PctInspectViz}{16.2}
\newcommand{\NDomainKnowledge}{56}
\newcommand{\PctDomainKnowledge}{11.1}
\newcommand{\NCheckReference}{45}
\newcommand{\PctCheckReference}{8.9}
\newcommand{\NBenchmark}{38}
\newcommand{\PctBenchmark}{7.5}
\newcommand{\NUnitTests}{14}
\newcommand{\PctUnitTests}{2.8}
\newcommand{\NAskAI}{14}
\newcommand{\PctAskAI}{2.8}
\newcommand{\NColleagueReview}{10}
\newcommand{\PctColleagueReview}{2.0}
\newcommand{\NMathByHand}{6}
\newcommand{\PctMathByHand}{1.2}
\newcommand{\NCrossCheckAI}{2}
\newcommand{\PctCrossCheckAI}{0.4}
\newcommand{\PctUnitTestsWhole}{3}
\newcommand{\PctColleagueReviewWhole}{2}

\newcommand{\NDhRunOnly}{26}

\newcommand{\NDbgMulti}{33}
\newcommand{\NDbgRunCode}{57}
\newcommand{\NDbgRunOnly}{31}
\newcommand{\NMscMulti}{34}
\newcommand{\NMscRunCode}{31}
\newcommand{\NMscRunOnly}{7}
\newcommand{\NAnaMulti}{28}
\newcommand{\NAnaRunCode}{24}
\newcommand{\NAnaRunOnly}{7}
\newcommand{\NVizInspectViz}{56}

\newcommand{\NAnaInspectOutput}{18}

\newcommand{\NExpUseCaseTests}{463}
\newcommand{\MeanLogDebug}{0.82}
\newcommand{\MeanLogNoDebug}{0.89}
\newcommand{\YearsDebug}{5.6}
\newcommand{\YearsNoDebug}{6.8}
\newcommand{\TDebugExp}{1.93}
\newcommand{\DfDebugExp}{130.8}
\newcommand{\PDebugExp}{.056}
\newcommand{\PBhDebugExp}{.28}
\newcommand{\NExpStrategyTests}{460}
\newcommand{\TReadExp}{1.58}
\newcommand{\DfReadExp}{147.9}
\newcommand{\PReadExp}{.12}
\newcommand{\PBhReadExp}{.58}
\newcommand{\DfStrategyCountExp}{458}
\newcommand{\RStrategyCountExp}{-.03}
\newcommand{\PStrategyCountExp}{.49}
\newcommand{\ChisqMscDf}{4}
\newcommand{\ChisqMsc}{13.33}
\newcommand{\PMscArea}{.010}
\newcommand{\PBhMscArea}{.046}
\newcommand{\PctMscEngineering}{21}
\newcommand{\PctMscLifeSci}{14}
\newcommand{\PctMscPhysSci}{14}
\newcommand{\PctMscSocialSci}{8}
\newcommand{\PctMscOther}{6}
\newcommand{\ChisqAnaDf}{4}
\newcommand{\ChisqAna}{11.86}
\newcommand{\PAnaArea}{.018}
\newcommand{\PBhAnaArea}{.046}
\newcommand{\PctAnaEngineering}{4}
\newcommand{\PctAnaLifeSci}{14}
\newcommand{\PctAnaPhysSci}{6}

\newcommand{\PctAnaOther}{17}
\newcommand{\NSoloExp}{513}
\newcommand{\RSoloExp}{.28}
\newcommand{\PSoloExp}{8\times10^{-11}}
\newcommand{\NGenaiExp}{512}
\newcommand{\RGenaiExp}{-.01}
\newcommand{\PGenaiExp}{.85}
\newcommand{\NEvalExp}{513}
\newcommand{\REvalExp}{.17}
\newcommand{\PEvalExp}{9\times10^{-5}}
\newcommand{\NGapExp}{512}
\newcommand{\RGapExp}{.21}
\newcommand{\PGapExp}{9\times10^{-7}}
\newcommand{\GapCrossLog}{0.68}
\newcommand{\GapCrossYears}{3.7}
\newcommand{\REvalSolo}{.30}
\newcommand{\PEvalSolo}{4\times10^{-12}}
\newcommand{\REvalGenai}{.33}
\newcommand{\PEvalGenai}{1\times10^{-14}}
\newcommand{\RSoloGenai}{.02}
\newcommand{\PSoloGenai}{.60}
\newcommand{\FUcEvalDfNum}{5}
\newcommand{\FUcEvalDfDen}{455}
\newcommand{\FUcEval}{3.53}
\newcommand{\PUcEval}{.004}
\newcommand{\DeltaRsqUcEval}{.036}
\newcommand{\FUcSoloDfNum}{5}
\newcommand{\FUcSoloDfDen}{455}
\newcommand{\FUcSolo}{1.38}
\newcommand{\PUcSolo}{.23}
\newcommand{\FUcGenaiDfNum}{5}
\newcommand{\FUcGenaiDfDen}{454}
\newcommand{\FUcGenai}{0.32}
\newcommand{\PUcGenai}{.90}
\newcommand{\NAncovaFig}{462}
\newcommand{\DfAncova}{459}
\newcommand{\MeanEvalDebug}{3.86}
\newcommand{\MeanEvalNoDebug}{4.19}
\newcommand{\BEvalDebug}{-0.30}
\newcommand{\TEvalDebug}{-3.12}
\newcommand{\PEvalDebugAnc}{.002}
\newcommand{\PBhEvalDebugAnc}{.010}
\newcommand{\MeanEvalViz}{4.32}
\newcommand{\MeanEvalNoViz}{4.07}
\newcommand{\BEvalViz}{0.24}
\newcommand{\TEvalViz}{2.59}
\newcommand{\PEvalVizAnc}{.010}
\newcommand{\PBhEvalVizAnc}{.025}
\newcommand{\FVmEvalDfNum}{5}
\newcommand{\FVmEvalDfDen}{452}
\newcommand{\FVmEval}{1.55}
\newcommand{\PVmEval}{.17}
\newcommand{\MeanEvalWithViz}{4.30}
\newcommand{\MeanEvalNoStratViz}{4.09}
\newcommand{\TEvalStratViz}{2.19}
\newcommand{\DfEvalStratViz}{120.1}
\newcommand{\PEvalStratViz}{.031}
\newcommand{\PBhEvalStratViz}{.10}
\newcommand{\MeanEvalWithRead}{4.27}
\newcommand{\MeanEvalNoStratRead}{4.09}
\newcommand{\TEvalStratRead}{2.06}
\newcommand{\DfEvalStratRead}{175.5}
\newcommand{\PEvalStratRead}{.041}

\newcommand{\DfStrategyCountEval}{457}
\newcommand{\RStrategyCountEval}{.05}
\newcommand{\PStrategyCountEval}{.29}
\newcommand{\BStrategyCountEval}{0.05}
\newcommand{\PStrategyCountEvalLm}{.23}
\newcommand{\NParentCollected}{1{,}272}
\newcommand{\NParentAnalytic}{868}
\newcommand{\NOpenCoding}{106}
\newcommand{\PctOpenCoding}{20}
\newcommand{\NPilotRound}{25}
\newcommand{\NSecondRound}{25}
\newcommand{\KAlphaPairwise}{0.76}
\newcommand{\KAlphaOverall}{0.70}
\newcommand{\NConsensusCoded}{50}
\newcommand{\NFlagged}{16}

\begin{document}

\ifarxiv
\title{How Researchers Use and Verify AI Coding Assistants: Tasks and
Validation Practices in Scientific Programming}
\fi

\author{Gabrielle O'Brien}
\email{elleobri@umich.edu}
\orcid{0009-0001-3198-3586}
\affiliation{%
  \institution{University of Michigan}
  \city{Ann Arbor}
  \state{Michigan}
  \country{USA}
}

\author{Reed Milewicz}
\email{rmilewi@sandia.gov}
\orcid{0000-0002-1701-0008}
\affiliation{%
  \institution{Sandia National Laboratories}
  \city{Albuquerque}
  \state{New Mexico}
  \country{USA}
}

\author{Nasir Eisty}
\email{neisty@utk.edu}
\orcid{0000-0001-5228-4664}
\affiliation{%
  \institution{University of Tennessee, Knoxville}
  \city{Knoxville}
  \state{Tennessee}
  \country{USA}
}

\renewcommand{\shortauthors}{\blinded{Anonymous authors}{O'Brien et al.}}

\begin{abstract}
\ifarxiv
Generative AI has entered research programming, yet there is little evidence
about which tasks researchers hand to it or how they decide whether its code
is correct. We draw on \NAnnotated{} free-text responses to a 2025 survey of
researchers who write code, most of them at U.S.\ universities. In each
response, a researcher recounts a single task from their own work, the way
they used an AI tool for it, and what they did to assess the result.
\fi
We coded the task and the evaluation
strategies reported, and related both to programming experience,
research area, and confidence ratings. Use was concentrated in five tasks:
data handling, visualization, debugging, mathematical/scientific computing,
and statistical analysis. Evaluation was informal and individual. Over half
of accounts described running the generated code, while automated tests and
review by another person were rare. Use cases and evaluation strategies
varied little with programming experience, but confidence did: Less
experienced programmers trusted the AI more than themselves, and experienced
programmers the reverse. Evaluation confidence was not associated with the strategies reported. Its
strongest correlates were confidence in the tool and in oneself. Validating AI contributions to scientific code rested
largely on individual judgment, outside shared infrastructure for testing or
review. Interfaces could support task-appropriate evaluation rather than
leave it to the user.
\end{abstract}

\ccsdesc[500]{Human-centered computing~Empirical studies in HCI}
\ccsdesc[300]{Software and its engineering~Software verification and validation}

\keywords{generative AI, research software, scientific programming, code generation, verification, validation, survey, content analysis}

\maketitle

\section{Introduction}
\label{sec:intro}

Modern science runs on software, and much of that software is written by
scientists rather than by software engineers
\citep{Kelly2007ASoftwareChasm,Segal2007SomeDevelopers}.
A crisis of productivity and credibility in research software in the 2000s
\citep{Storer2017BridgingChasm,johanson2018software} brought training
programs \citep{carpentries,intersect_training}, open science policies \citep{mckiernan2016open},
and a research software engineering workforce
\citep{Baxter2012ResearchSoftwareEngineer,cohen2020four}. Even so, much
scientific programming is still done by what \citet{Segal2007SomeDevelopers}
called ``professional end-user developers'': domain experts who program in
service of their science. Knowing whether code meets its scientific
requirements is hard, even for expert developers of research software
\citep{Kanewala2014TestingReview,Sanders2008DealingWithRisk,Kelly2015ScientificSoftware}.
And the cost of not knowing can be high: coding errors that slipped through
have forced retractions of published findings in
healthcare \citep{JCO2016Retraction} and public policy
\citep{Karraker2015AuthorsExplanation}, among other fields
\citep{Kanewala2014TestingReview}.

Into this setting have arrived tools that generate working code from
natural language. Scientists and research software engineers (RSEs) have adopted
them quickly
\citep{VanTuyl2025StateOfAI,Chugunova2026WhoUses,OBrien2025HowProgram},
and the tools are increasingly capable of taking on more of the work, from
inline completion and conversational chatbots to agents that execute code
and iterate on their own \citep{Chen2026CodeWithMe}.

Deciding how and when to rely on an automated system is a long-standing problem.
Research on automation holds that trust should match what the system can
actually do \citep{LeeSee2004TrustAutomation}, but in practice people may tend to over-rely
on automated aids 
\citep{Goddard2012AutomationBias}. Among knowledge workers, for example, confidence in
a generative AI tool is associated with less scrutiny of its output \citep{Lee2025TheWorkers}.
Scientific programmers often write code to explore data or simulate systems
that cannot be observed directly, so they do not already know what the
correct result should look like. Now they must also contend with generated
code that may be plausible but wrong. Furthermore, models signal their own uncertainty poorly
\citep{Spiess2025CalibrationCode}, so judging the output becomes the user's
job (and one that can consume a large share of developers' time
\citep{Mozannar2024ReadingProgramming,Vaithilingam2022ExpectationModels}). 

In professional software engineering, automated tests and structured code review exist to detect issues with programs that may escape an individual's notice. 
In the scientific community, where verification is rarely standardized or
supported by shared infrastructure
\citep{Carver2022AStates,Hannay2009HowSoftware}, deciding whether generated code is acceptable may often be a matter of individual interactions with their AI tools of choice. 

How programmers work with AI tools has been studied through laboratory
studies of developers and students
\citep{Barke2023GroundedModels,Vaithilingam2022ExpectationModels,Mozannar2024ReadingProgramming,Prather2023ItsWeird,Nguyen2024HowBeginning,Tang2024AActions}, surveys of
developers \citep{Liang2024ALargeScale}, and observations of professional
data analysts \citep{Gu2024HowAnalyses}.
But the same qualities that make scientific programming distinctive make it
unclear how far those findings generalize. Studies of scientists who program
are still sparse, drawn from the more professionalized research software
engineering community or from a small number of open-ended survey responses
\citep{OBrien2025HowProgram,VanTuyl2025StateOfAI,Besser2026HowGenerative}
(Section~\ref{sec:related}).

Here, we examine \NAnnotated{} written \emph{accounts} of generative AI
use in research programming, asking not only what scientists delegate to
these tools but how they judge the result and how confident they are in
that judgment. An account is one respondent's answers to a set of
open-ended prompts about a single episode of AI use, together with the
confidence ratings attached to it. The accounts were collected during a 2025 cross-institutional survey of
scientists who program; the survey's descriptive results are published
separately \blinded{[withheld for review]}{\citep{OBrien2025AProgram}}, and this paper is the first to analyze
the open-ended section. Each respondent
described one specific, recent task for which they used their primary AI tool, how they used it, and how they determined whether the output was
acceptable, using prompts adapted from \citeauthor{Lee2025TheWorkers}'s survey of
knowledge workers \citep{Lee2025TheWorkers}. Respondents also rated their confidence in doing the
task alone, in the tool's ability to do it, and in their own evaluation of
the output. We coded each account along two dimensions---the scientific
task motivating the request and the strategies reported for evaluating the
result---and related those codes to programming experience, research
area, and the confidence ratings. Throughout, we analyze what respondents
\emph{say they do} in a single recalled episode, which may not match their
exact behavior. We ask:

\begin{itemize}
  \item \textbf{RQ1.} What tasks do scientists who program delegate to
  generative AI, and how do they report evaluating the output?
  \item \textbf{RQ2.} Do reported use cases and evaluation strategies vary
  with programming experience or research area?
  \item \textbf{RQ3.} How is confidence---in the tool, in oneself, and in
  one's evaluation of the output---related to experience, to the task, and
  to the evaluation strategies reported?
\end{itemize}

Our main findings are as follows:

\begin{enumerate}
  \item Use is concentrated in five tasks---data handling, visualization,
  debugging, mathematical/scientific computing, and statistical analysis---which together account for \PctTopFiveUseCaseShare\% of accounts that
  received a use-case code (Section~\ref{sec:qual}).
  \item Reported evaluation is informal and individual. Over half of
  accounts describe running the generated code (\PctRunCode\%), followed by
  inspecting its output, reading the code, and inspecting visualizations.
  Automated tests are mentioned in \PctUnitTestsWhole\% of accounts and review
  by another person in \PctColleagueReviewWhole\% (Section~\ref{sec:qual}).
  \item Neither use cases nor evaluation strategies vary much with
  programming experience. The few detectable differences are by research
  area, in expected directions (Section~\ref{sec:quant}).
  \item Confidence does vary with experience. For the episode they
  described, less experienced programmers rated the tool above themselves,
  while more experienced programmers rated themselves above the tool
  (Section~\ref{sec:quant}).
  \item Respondents' confidence in their evaluation is not reliably
  associated with the evaluation strategies they reported. Evaluation
  confidence is more strongly correlated with their confidence in the tool and in their own
  ability to do the task alone (Section~\ref{sec:quant}).
\end{enumerate}

Together, these findings suggest that in the episodes respondents
described, validating AI contributions to scientific code rested largely on
individual judgment, exercised outside shared infrastructure for testing or
review. We discuss what this implies for tools that could support
verification where correctness is not visible in the output, for training
that makes experienced scientists' informal checks explicit, and for the
integrity of research that increasingly depends on machine-generated code
(Section~\ref{sec:discussion}).

\section{Related Work}
\label{sec:related}

\subsection{Programming with AI code generation tools}

HCI and software engineering researchers have examined how programmers
work with tools that generate code from natural language, which
\citet{Sarkar2022WhatIsIt} argue is a distinct activity from conventional
programming. \citet{Vaithilingam2022ExpectationModels}
found that most participants preferred GitHub Copilot to conventional
autocomplete but struggled to understand, edit, and debug the longer blocks
of code it produced. \citet{Barke2023GroundedModels} developed a grounded
theory of Copilot use in which programmers alternate between an
\emph{acceleration} mode, where the tool completes code the programmer
already had in mind, and an \emph{exploration} mode, where it is used to
discover approaches. Evaluation strategies differed between the two modes.
\citet{Mozannar2024ReadingProgramming} modeled where developers' time goes
during Copilot sessions, finding that a substantial share is spent verifying
suggestions rather than writing code, and \citet{Tang2024AActions} used eye
tracking and IDE logs to characterize how developers validate and repair
generated code. Beginners appear to struggle most: \citet{Zi2025ICode}
found that novices had difficulty understanding code generated by LLMs
even when it was correct. Outside the laboratory, \citet{Liang2024ALargeScale}
surveyed 410 developers about their use of AI programming assistants,
finding that the most common motivations were reducing keystrokes and
finishing tasks faster, while the most common reasons for not using
suggestions were that the generated code did not meet requirements or that
developers lacked control over the output. Evidence on productivity is
mixed: A controlled experiment found that developers completed a task
substantially faster with Copilot \citep{Peng2023TheCopilot}, while a field
experiment with experienced open-source developers found that AI tools
slowed them down \citep{Becker2025MeasuringProductivity}.

\subsection{Verifying AI output and calibrating reliance}
\label{sec:rw-reliance}

Whether to accept an AI output is a reliance decision. Research on trust in
automation treats calibrated trust, in which reliance tracks the system's
actual reliability, as the goal rather than maximal trust
\citep{LeeSee2004TrustAutomation}. Overreliance on decision support is well
documented across domains \citep{Goddard2012AutomationBias}. Experiments
with AI decision aids find that explanations can raise reliance whether or not
the AI is right \citep{Bansal2021DoesTheWhole}, and that forcing users to engage
with the task before seeing the AI's answer reduces overreliance at a cost in
effort \citep{Bucinca2021ToTrust}. People also rely on an AI more when
checking its output themselves is costly \citep{Vasconcelos2023Explanations}.
Users may also be poorly placed to judge that reliability, because they
overestimate what language models know \citep{Steyvers2025WhatLLMsKnow}.
Among software developers, trust in code generation tools rests on the
tool's perceived ability, integrity, and benevolence and varies with the
context of use \citep{Wang2024InvestigatingTrust}, and is shaped by the experiences
peers share \citep{Cheng2024ItWouldWork}.

A parallel literature examines how users check AI-generated data
analyses. \citet{Gu2024HowAnalyses} observed
22 professional analysts verifying AI-generated analyses and found that they
began with procedure-oriented checks (what did the tool do?) and shifted to
data-oriented checks (does the result make sense?) once something looked
off. Studies of end-user programmers disagree about how well such checks
work: Inspecting the shape and contents of data objects by eye has been reported
both as a successful check \citep{Gu2024HowAnalyses} and as a route to
overconfidence in incorrect results \citep{Ragavan2022GridBook}, a pattern
with a long history in end-user programming
\citep{Panko2008TwoDevelopment}. In a survey of 319 knowledge workers,
\citet{Lee2025TheWorkers} found the same asymmetry between confidence in
the tool and confidence in oneself, and that generative AI shifts critical
effort toward verifying and integrating the tool's output rather than
producing one's own. \citet{Storey2026FromAI}
argue that code produced without the programmer's full understanding accrues
a distinct kind of debt in comprehension and intent.

Novices and end-user programmers appear especially exposed.
\citet{Prather2023ItsWeird} observed beginners accepting generated code
without validating it, and \citet{Nguyen2024HowBeginning} and
\citet{Liu2023WhatItWants} found that non-experts struggle to steer code
generation through prompts. These findings are consistent with the long-standing habit of end-user
programmers to tweak code they do not fully understand until its output
looks right \citep{Lau2021TweakIt}. \citet{Tankelevitch2024Metacognitive}
frame deciding when and how thoroughly to check as a metacognitive demand
these tools place on the user. Interface cues that flag likely errors can direct programmers' attention
to them, but highlighting tokens by generation probability, as commercial
tools do, did not change behavior in one study
\citep{Vasconcelos2024GenerationProbabilities}.

\subsection{Scientists who program}
\label{sec:rw-research-software}

Much research software is written by scientists themselves. In a 2014
survey of 417 researchers at UK universities, 56\% developed their own
software, and a fifth of those had no training in software development
\citep{Hettrick2014Survey}. These scientists write code to do science
rather than to build software
\citep{Howison2011ScientificSoftwareProduction}, and their practices differ
from industrial software engineering for reasons that are often legitimate.
They code to explore data and ideas rather than to build a product
\citep{SutherlandKeller2025ResearchSoftware}, requirements are discovered
as the software evolves alongside the science
\citep{Segal2008ModelsDevelopment,Kelly2015ScientificSoftware}, and testing
is informal or absent by conventional measures
\citep{eisty2025testing,Carver2022AStates}, although many scientists hold a
broader notion of verification and validation that covers the mathematics
their code implements and the physical experiments it models
\citep{oberkampf2010verification}. Much of their code lives in scripts and
computational notebooks, where it is run in short interactive fragments
rather than strictly as written
\citep{Kery2017ExploringProgramming,Head2019ManagingMesses}, and testing
may happen through expertly curated diagnostic plots rather than a battery
of tests \citep{Paine2017WhoHasPlots}. Research software engineers (RSEs), a
professional workforce that builds and maintains research software as its
job \citep{Baxter2012ResearchSoftwareEngineer,cohen2020four}, adopt more
conventional practices, although even among them testing and review are far
from universal \citep{Carver2022AStates}. It is the scientists writing
their own code, not RSEs, who concern us here. Tools for this population
have to fit its needs and values rather than import industrial practice
wholesale. This paper asks how generative AI tools are being fitted into
these practices: what scientists delegate to them, and how they come to
trust the output.

Evidence that scientists have adopted these tools is accumulating. In a 2024 survey of over 6,000 researchers at two large German
research organizations, writing code was among the two most common uses of
AI, reported by 43.2\% of respondents \citep{Chugunova2026WhoUses}. In a
2026 survey of faculty and staff who actively develop or maintain research
software at one U.S.\ university, about a third reported using generative AI
tools \citep{Besser2026HowGenerative}. Surveying the research software
engineering community, \citet{VanTuyl2025StateOfAI} found that roughly four
in five reported using AI tools on the job. Why programmers at
different experience levels use these tools --- to attempt work they
could not do alone, or to speed up work they could --- is not yet clear
\citep{Lee2025TheWorkers}.

Beyond adoption rates, less is known about how these tools are used in
practice. Interviews with 14 scientists who programmed with AI assistance
suggest that the tools often function as a substitute for documentation, and that
verification is largely informal: running code and inspecting the output,
reading line by line, or asking the tool to explain itself
\citep{OBrien2025HowProgram}. Two surveys have since coded free-text
accounts of AI use in research software: Among research software engineers
and adjacent staff, common themes were clarifying tasks and language, code
generation, refactoring, and retrieving reference information for
undocumented functions \citep{VanTuyl2025StateOfAI}. Among developers at one
university, the themes were scaffolding, debugging, natural-language data transformation,
and cognitive offloading \citep[from at most 24
responses]{Besser2026HowGenerative}.
The present study adds a larger,
cross-disciplinary corpus in which each account is coded for both the task
and the strategy used to evaluate the output.

\section{Methods}

\subsection{Survey}

We report a mixed-methods study: a qualitative content analysis of written
accounts, followed by quantitative analysis relating the resulting codes to
respondents' experience, research area, and confidence ratings. The data analyzed here was collected during a 2025 survey of research
scientists who program. The full survey instrument and several descriptive results
(adoption rate, tool preferences, and perceived productivity) are reported in a
separate publication \blinded{[withheld for review]}{\citep{OBrien2025AProgram}}; the open-ended accounts analyzed
here have not been reported before.
Briefly, the instrument comprised seven sections administered in Qualtrics
(estimated 10--15 minutes), covering demographics, programming background,
organizational coding practices, generative AI tool experience, perceived
productivity, an open-ended use-case section, and reasons for non-adoption
(if applicable). Recruitment was through mailing lists and online
communities for scientific programming (e.g., the US Research Software
Engineering association, pyOpenSci, and a targeted list at the authors'
institution\blinded{ [withheld for review]}{, the University of Michigan}).
Because the link was shareable, the response rate is unknown. The study
was approved by the authors' institutional review board\blinded{ [withheld for review]}{ at the University of Michigan},
 responses were anonymous,
and no IP addresses were collected. Of \NParentCollected{} responses collected
between July 10 and August 25, 2025, we excluded incomplete responses, those
failing consent or eligibility screens, and respondents who never program in
their research, leaving \NParentAnalytic{} in the survey's analytic sample.

This analysis concerns the open-ended use-case section, shown only to
respondents who reported using a generative AI tool in their
research-related programming (those who had never tried such tools or reported that they had given up were routed to a non-adoption branch of the
survey instead). After selecting their primary tool, eligible respondents
answered three open-ended prompts. The format of this section, in which a
respondent describes one specific recent example of AI use and then rates
their confidence about it, was adapted from a survey of knowledge workers
by \citet{Lee2025TheWorkers}; we reworded the items for research
programming.

The prompts were:

\begin{itemize}
  \item \emph{Use case:} ``Think about one specific, real-world example of
  how you used your primary generative AI tool while doing research-related
  programming. What were you trying to achieve?''
  \item \emph{Tool use:} ``How did you use the tool in this example? If
  possible, please include any prompts\ldots''
  \item \emph{Evaluation:} ``How did you determine if the output of the
  generative AI tool was acceptable?''
\end{itemize}

Respondents also rated three confidence items on a 5-point scale
(1 = ``not at all confident,'' 5 = ``extremely confident''). These are the
three confidence constructs from \citet{Lee2025TheWorkers}: confidence in
doing the task without generative AI, confidence in the tool's ability to do
it, and confidence in one's own ability to evaluate the output. That study
found that the first and second were associated in opposite directions with
critical evaluation of AI output, which motivated retaining all three here.

\begin{itemize}
  \item \emph{Solo confidence:} confidence in doing the task without
  generative AI
  \item \emph{GenAI confidence:} confidence in the tool's ability to do the
  task
  \item \emph{Evaluation confidence:} confidence in evaluating the tool's
  output in the course of normal work
\end{itemize}

There were \NAnnotated{} accounts from the use-case section, which were the
target of our qualitative analysis.

\subsection{Qualitative analysis}

We conducted a qualitative content analysis of open-ended survey responses
describing how researchers used generative AI tools for programming tasks.
We developed two parallel codebooks: a use-case codebook capturing the
primary programming task for which the respondent used a generative AI tool (e.g., debugging, data handling, visualization), and an evaluation
codebook capturing how respondents verified or validated AI-generated output (e.g., running code, inspecting output, consulting reference
documentation).

To generate the codebooks, the first author first conducted a round of open coding of \NOpenCoding{} randomly selected (\PctOpenCoding\%) accounts and drafted each codebook from that round. Using the draft codebooks, all three authors
conducted a round of coding on a new sample of \NPilotRound{} responses,
then met to discuss disagreements. These discussions led to refinements in code definitions (for example, the category ``Systems \&
Hardware'' was expanded to include creating code to interact with high-performance computing systems, and ``Code comprehension'' was clarified to
refer only to trying to understand code written by another person and not an
AI tool). We also clarified rules for interpreting common ambiguous words. The word ``test,'' for example, occurred frequently in contexts such as ``I tested the suggestion'' or ``it compiled and worked when I tested it.''
``Testing'' can have a specific meaning in software development, referring
to a ``harness'' of checks that are run after changes to the codebase (often
with some automation). Based on prior research showing that this form of testing is infrequent in scientific software development \citep{Carver2022AStates,eisty2025testing} and our best judgment of the contexts in which the term ``test'' typically occurred in
survey responses, we decided that phrases such as ``I tested the suggestion''
would be labeled ``Run code'', referring to manually initiating code
execution. We applied the ``test suite'' label only when the account specifically indicated use of a test harness (such as ``unit tests'' or ``integration testing'').

After this discussion, we also identified two interpretive rules: first, code assignments must be grounded in explicit textual
evidence. In practice, this means that if a person describes making a plot
with AI assistance but reports only that they looked at the resulting plot, we applied only the label ``Inspect visualization,'' even though having a visual artifact to inspect implies that they must have executed the generated code. We restricted ourselves to a content analysis of what respondents reported as their evaluation strategy, since we could not observe the steps they actually took. A second
interpretive rule is that the use case should be coded according to the primary reason the respondent describes using the AI tool. For example,
if a respondent indicates that they used ChatGPT to help reshape a dataframe from long to wide format but had to troubleshoot its suggestions briefly, the account would be coded only as ``Data handling,'' not ``Debugging.''

We then conducted a second round of independent coding with another
\NSecondRound{} randomly selected, previously uncoded accounts. We met to discuss codes and made a few minor refinements to the codebook (for example, adding a new code ``Validation activities'' for use cases involving validating the correctness of an existing codebase). After these refinements, the median pairwise Krippendorff's alpha with MASI (Measuring Agreement on Set-valued Items) distance was $\alpha = \KAlphaPairwise$. The overall agreement across all three raters was $\alpha = \KAlphaOverall$. At this point, all three authors had coded \NConsensusCoded{} accounts and we had reached a consensus on the codebook definitions. The first author annotated the remaining accounts with the finalized codebooks and applied them retroactively to the accounts from the initial open-coding round.
During this round, we flagged \NFlagged{} accounts for discussion because of ambiguous language, and all three raters met again to resolve their codes by consensus.

Not every response could be coded. Some respondents skipped questions in
this survey section, and others gave answers too vague to label confidently
(describing a use case as simply ``coding''). We did not apply codes when we judged that too little information was present.

\subsection{Quantitative analysis}

We follow several statistical conventions for our quantitative analyses.
Programming experience is measured in years and analyzed as
$\log_{10}(\text{years} + 1)$. Where helpful, we back-transformed to years for interpretation. Confidence items are 1--5 ratings. All two-group comparisons
are Welch's $t$-tests, which do not assume equal variances (group sizes here
are often very unequal). Within each family of per-code tests we control the
false discovery rate with the Benjamini--Hochberg (BH) procedure and report
both uncorrected and BH-adjusted $p$-values. For statistical power, per-code
analyses are restricted to the five most frequent codes in each family (use
cases: debugging, data handling, visualization, mathematical/scientific
computing, statistical analysis; evaluation strategies: run code, inspect
output, read code, inspect visualization, domain knowledge/intuition).
Analyses were conducted in R. Quantitative analysis scripts were written
with the assistance of Claude Code. The authors specified the statistical
tests and chart types, and used Claude Code for assistance in
implementation.

\section{Results}

Of \NAnnotated{} use-case accounts annotated, \NDropped{} (\PctDropped\%)
could not be assigned codes from either codebook and were excluded for data quality, leaving an analytic sample of $N = \NAnalytic$, that is, every account that received at least one use-case or evaluation code.
Within this sample, \NUseCaseCoded{} accounts received at least one
use-case code and \NValCoded{} at least one evaluation code. The subsets differ because a few accounts described only a use case or only an evaluation strategy. Some accounts carried more than one label per
codebook, as respondents sometimes described multiple use scenarios or
evaluation strategies.

Before describing major themes from the qualitative analysis, we briefly
summarize the demographics of respondents to contextualize their accounts.

\subsection*{Demographic information}

Table~\ref{tab:demographics} summarizes the analytic sample. Respondents were
concentrated in U.S. higher education, with \PctUS\% based in the United
States and \PctHigherEd\% at higher-education institutions. Men somewhat
outnumbered women (\PctMen\% to \PctWomen\%). Most respondents were
early-career: student research assistants were the largest group
(\PctStudentRA\%), and faculty made up \PctFaculty\%. The most-represented
research areas were the life sciences (\PctLifeSci\%), engineering
(\PctEngineering\%), the social sciences (\PctSocialSci\%), and the physical
sciences (\PctPhysSci\%).

\begin{table}[htbp]
\centering
\small
\caption{Respondent characteristics for the analytic sample ($N = 506$). Percentages use $N$ as the denominator except where noted. Languages were multi-select.}
\label{tab:demographics}
\begin{tabular}[t]{l r r}
\toprule
Characteristic & $n$ & \% \\
\midrule
\addlinespace
\multicolumn{3}{l}{\textit{Country}} \\
\quad United States & 494 & 97.6 \\
\quad Other country & 9 & 1.8 \\
\quad Not reported & 3 & 0.6 \\
\addlinespace
\multicolumn{3}{l}{\textit{Organization}} \\
\quad Higher education & 474 & 93.7 \\
\quad National laboratory & 12 & 2.4 \\
\quad Private sector & 6 & 1.2 \\
\quad Other & 14 & 2.8 \\
\addlinespace
\multicolumn{3}{l}{\textit{Gender}} \\
\quad Man & 271 & 53.6 \\
\quad Woman & 213 & 42.1 \\
\quad Non-binary or gender-diverse & 13 & 2.6 \\
\quad Prefer not to say & 8 & 1.6 \\
\quad Not reported & 1 & 0.2 \\
\addlinespace
\multicolumn{3}{l}{\textit{Position}} \\
\quad Student research assistant & 205 & 40.5 \\
\quad Research staff & 125 & 24.7 \\
\quad Faculty & 84 & 16.6 \\
\quad Post-doctoral researcher & 76 & 15.0 \\
\quad Research software engineer & 10 & 2.0 \\
\quad Other & 6 & 1.2 \\
\addlinespace
\multicolumn{3}{l}{\textit{Research area}} \\
\quad Life sciences & 145 & 28.7 \\
\quad Engineering & 104 & 20.6 \\
\quad Social sciences & 54 & 10.7 \\
\quad Physical sciences & 52 & 10.3 \\
\quad Computer and information sciences & 44 & 8.7 \\
\quad Psychology & 26 & 5.1 \\
\quad Mathematics and statistics & 22 & 4.3 \\
\quad Geosciences and ocean sciences & 16 & 3.2 \\
\quad Other & 43 & 8.5 \\
\bottomrule
\end{tabular}
\hfill
\begin{tabular}[t]{l r r}
\toprule
Characteristic & $n$ & \% \\
\midrule
\addlinespace
\multicolumn{3}{l}{\textit{Programming frequency}} \\
\quad Daily & 254 & 50.2 \\
\quad Weekly & 173 & 34.2 \\
\quad Monthly & 45 & 8.9 \\
\quad Less than once a month & 34 & 6.7 \\
\addlinespace
\multicolumn{3}{l}{\textit{Programming experience (years)}} \\
\quad Median (IQR) & \multicolumn{2}{l}{6 (4--10.5)} \\
\quad Not reported & 11 & 2.2 \\
\addlinespace
\multicolumn{3}{l}{\textit{Languages used (multi-select)}} \\
\quad Python & 356 & 70.4 \\
\quad R & 285 & 56.3 \\
\quad MATLAB & 156 & 30.8 \\
\quad Bash & 104 & 20.6 \\
\quad C++ & 76 & 15.0 \\
\addlinespace
\multicolumn{3}{l}{\textit{Primary generative AI tool}} \\
\quad ChatGPT & 312 & 61.7 \\
\quad GitHub Copilot & 59 & 11.7 \\
\quad Custom tool provided by organization & 29 & 5.7 \\
\quad Google Gemini & 28 & 5.5 \\
\quad Claude & 22 & 4.3 \\
\quad Cursor & 8 & 1.6 \\
\quad Microsoft Copilot & 8 & 1.6 \\
\quad Claude Code & 7 & 1.4 \\
\quad Perplexity & 6 & 1.2 \\
\quad Other & 16 & 3.2 \\
\quad Not reported & 11 & 2.2 \\
\bottomrule
\end{tabular}
\end{table}

\subsection*{Programming practices}

Respondents were active programmers. Median programming experience was \MedianYears{} years (IQR \IqrYearsLo--\IqrYearsHi, max \MaxYears), and \PctWeeklyPlus\% programmed at least weekly (\PctDaily\% daily, \PctWeekly\% weekly). Python (\PctPython\%) and R (\PctR\%) dominated language use, followed by MATLAB (\PctMatlab\%), Bash (\PctBash\%), and C++ (\PctCpp\%). The long tail covered Stata, JavaScript, C, Fortran, Java, Julia, and Rust.
By design, only survey respondents who indicated that they used a generative AI tool for programming were asked to provide a use case. Respondents overwhelmingly named ChatGPT as their primary tool (\PctChatGPT\%), followed by
GitHub Copilot (\PctCopilot\%), institution-provided custom tools
(\PctCustomTool\%), Google Gemini (\PctGemini\%), and Claude (\PctClaude\%,
not including Claude Code). Roughly half the sample (\PctVersionControl\%) reported using version control ``about half the time'' or more, while regular use of formal testing and review practices was substantially lower: code review (\PctCodeReview\%), unit tests (\PctUnitTestPractice\%), system tests (\PctSystemTestPractice\%), and regression tests (\PctRegressionTestPractice\%).

\subsection{Qualitative analysis}
\label{sec:qual}

\begin{figure}[htbp]
  \centering
  \includegraphics[width=\textwidth]{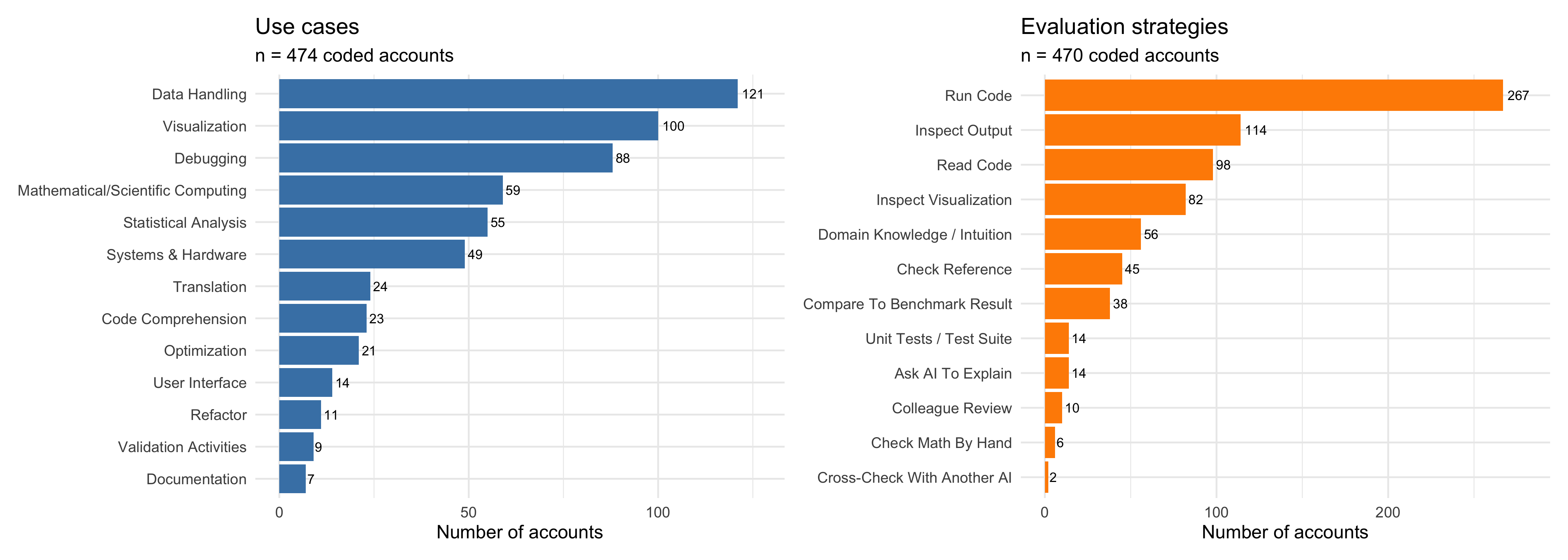}
  \Description{Two horizontal bar charts: counts of the thirteen use-case codes (left, data handling highest) and the twelve evaluation codes (right, run code highest).}
  \caption{Frequency of use-case codes (left; $n = \NUseCaseCoded$ accounts with
  at least one use-case code) and evaluation codes (right;
  $n = \NValCoded$ accounts with at least one evaluation code). Codes are not mutually exclusive, since an account may carry several codes from each codebook. Percentages in the text use the full analytic sample
  ($N = \NAnalytic$) as the denominator.}
  \label{fig:code-counts}
\end{figure}

We identified \NUseCaseTypes{} use cases for generative AI in research
programming (Figure~\ref{fig:code-counts}, left). The five most common use cases --- data
handling ($n=\NDataHandling$,
\PctDataHandling\%), visualization ($n=\NVisualization$,
\PctVisualization\%), debugging ($n=\NDebugging$, \PctDebugging\%),
mathematical/scientific computing ($n=\NMathSci$, \PctMathSci\%), and
statistical analysis ($n=\NAnalysisUC$, \PctAnalysisUC\%) --- appeared in about three-quarters of accounts that
received a use-case code, and so we focused our qualitative reporting on these five. The remaining eight
use cases (translation, code comprehension, optimization, user interface
work, refactoring, validation activities, documentation, and systems \&
hardware tasks) each appeared in a smaller share of accounts. Most responses
described a single use case (\PctSingleUseCase\%; mean \MeanUseCaseCodes{}
codes per account).

\begin{figure}[htbp]
  \centering
  \includegraphics[width=0.9\textwidth]{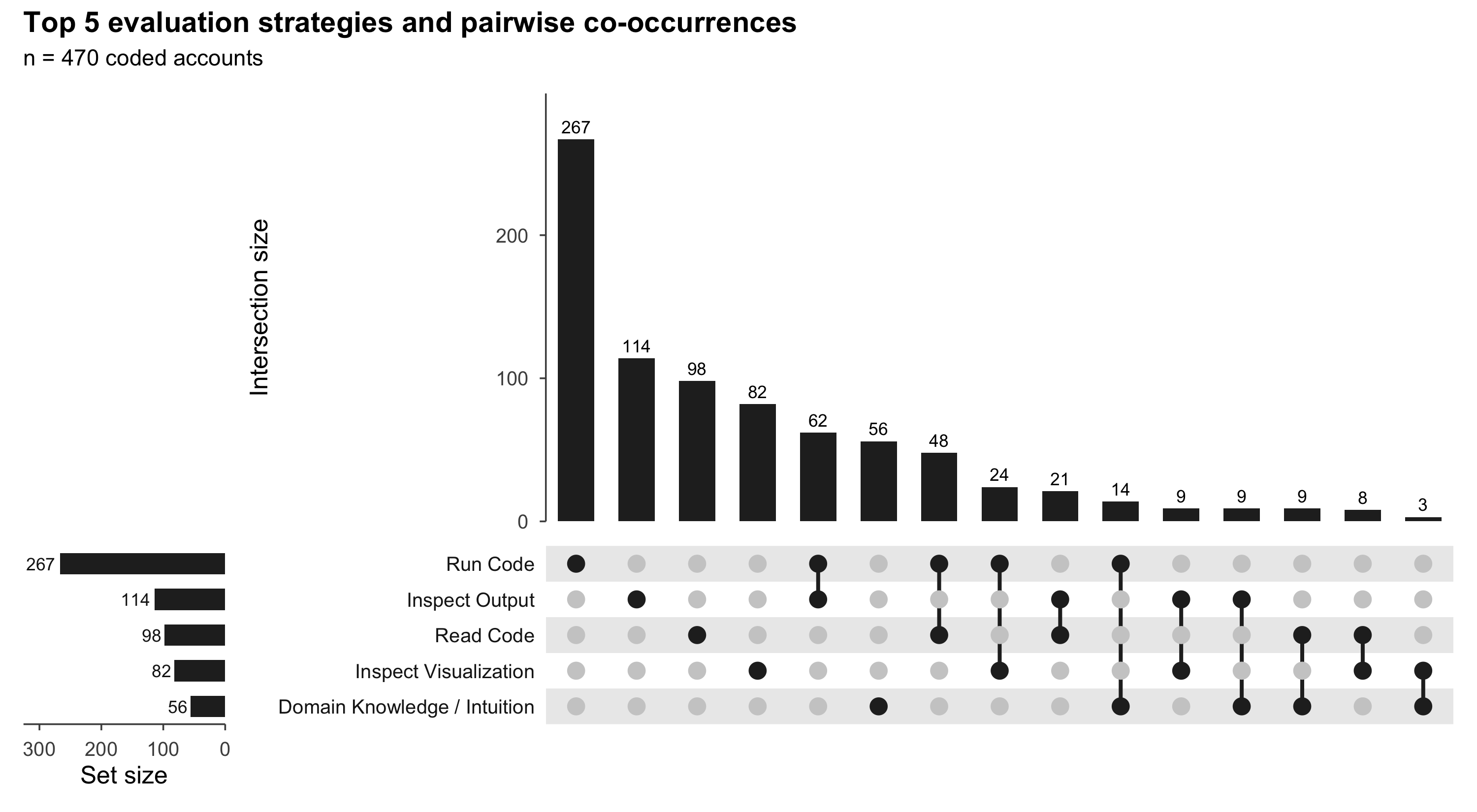}
  \Description{UpSet-style plot of the five most common evaluation strategies, with bars for each strategy alone and for each pair.}
  \caption{The five most common evaluation strategies and their pairwise
  co-occurrences. Horizontal bars (left) give each strategy's overall
  count; vertical bars give the number of accounts carrying the
  indicated single code or code pair (connected dots), counted regardless of any other codes also present. Unlike a conventional UpSet plot, columns are not mutually exclusive intersections. $n = \NValCoded$ accounts
  with at least one evaluation code.}
  \label{fig:upset}
\end{figure}

In parallel, we coded the strategies that respondents reported using to evaluate the AI's output in their given use case (Figure~\ref{fig:code-counts},
right). Use of multiple strategies was common: \PctOneStrategy\% of accounts
described a single evaluation strategy ($n=\NOneStrategy$), \PctTwoStrategies\% described two
($n=\NTwoStrategies$), and \PctThreePlusStrategies\% described three or more ($n=\NThreePlusStrategies$; mean \MeanValCodes{} codes per account). Pairwise combinations of the most common strategies are shown in Figure~\ref{fig:upset}. The modal strategy by a wide margin was running
the code ($n=\NRunCode$, \PctRunCode\% of accounts), followed by inspecting outputs
($n=\NInspectOutput$, \PctInspectOutput\%), reading the code ($n=\NReadCode$, \PctReadCode\%), inspecting a
visualization ($n=\NInspectViz$, \PctInspectViz\%), and drawing on domain knowledge or
intuition ($n=\NDomainKnowledge$, \PctDomainKnowledge\%).
Tables~\ref{tab:usecase-codebook} and~\ref{tab:eval-codebook} give the
definition, count, and an example account for every code in each codebook.

\begingroup
\footnotesize
\setlength{\LTcapwidth}{\textwidth}

\begin{longtable}{p{2.6cm} l p{5.0cm} p{5.7cm}}
\caption{Use-case codebook: the task motivating the request, coded from the primary reason that the respondent gave for using the AI tool. $n$ (\%) is accounts in the analytic sample ($N = \NAnalytic$) carrying each code; quotes are verbatim answers to the use-case prompt, reproduced as written.}
\label{tab:usecase-codebook}\\
\toprule
Code & $n$ (\%) & Definition & Example account \\
\midrule
\endfirsthead
\multicolumn{4}{l}{\tablename~\thetable{} (continued)}\\
\toprule
Code & $n$ (\%) & Definition & Example account \\
\midrule
\endhead
\bottomrule
\endlastfoot

Data handling & \NDataHandling{} (\PctDataHandling) &
Move, manipulate, process, merge, or reformat datasets, usually as steps that prepare data for later analysis. &
``doing a somewhat complicated join on two dataframes from different csvs'' \\
\addlinespace[2pt]
Visualization & \NVisualization{} (\PctVisualization) &
Create, modify, or format data visualizations, plots, graphs, or figures. &
``How to modify my code so that my Python script could go through a list of
colors and assign a different color to each bar in my bar chart plot'' \\
\addlinespace[2pt]
Debugging & \NDebugging{} (\PctDebugging) &
Troubleshoot why something did not work as expected; find and/or repair the
root cause of an error or unexpected behavior in code. &
``I was trying to run a logistic regression model and I was getting an
error'' \\
\addlinespace[2pt]
Mathematical/scientific computing & \NMathSci{} (\PctMathSci) &
Domain-specific algorithms, such as solving integrals or systems of equations, implementing formal logic, and building control systems or simulations. &
``find solution for a nonlinear equation system'' \\
\addlinespace[2pt]
Statistical analysis & \NAnalysisUC{} (\PctAnalysisUC) &
Implement methods for finding patterns in data, calculating statistics or
running statistical tests, and fitting models to data; includes signal processing techniques for pattern recognition. &
``Perform exploratory analysis on a gene expression matrix.'' \\
\addlinespace[2pt]
Systems \& hardware & \NSystemsHw{} (\PctSystemsHw) &
Interface with physical devices, handle device drivers, control instruments
for experiments; provision resources on HPC or cloud compute. &
``I needed boilerplate code for reading / writing to a binary file. I wanted
to use the c++ std lib rather than C utilities'' \\
\addlinespace[2pt]
Translation & \NTranslation{} (\PctTranslation) &
Convert code from one programming language (or library or framework) to
another. &
``convert a script I had written in python to MATLAB'' \\
\addlinespace[2pt]
Code comprehension & \NCodeComp{} (\PctCodeComp) &
Understand code written by another person (not AI-generated code). &
``Used AI to understand the code from a previously published paper that was
written in an unfamiliar language'' \\
\addlinespace[2pt]
Optimization & \NOptimization{} (\PctOptimization) &
Improve aspects of code performance, such as memory usage or speed. &
``Speed up Euclidean distance generation for an R package'' \\
\addlinespace[2pt]
User interface & \NUserInterface{} (\PctUserInterface) &
Create interactive elements, dashboards, or web applications; build
user-facing interfaces or interactive tools. &
``I wanted to create a webpage to view some data I annotated.'' \\
\addlinespace[2pt]
Refactor & \NRefactor{} (\PctRefactor) &
Reorganize or simplify existing code without changing its core functions,
usually to make code more readable or maintainable. &
``reorganize code for more flexibility and readability'' \\
\addlinespace[2pt]
Validation activities & \NValidationAct{} (\PctValidationAct) &
Validate the correctness of an existing code base, for example by
generating tests or checks for code the respondent already had. &
``Create unit tests to verify/validate a new code feature'' \\
\addlinespace[2pt]
Documentation & \NDocumentation{} (\PctDocumentation) &
Write comments, docstrings, or documentation for code. &
``I was trying to write documentation for some undocumented functions in an
R package'' \\
\end{longtable}

\begin{longtable}{p{2.6cm} l p{5.0cm} p{5.7cm}}
\caption{Evaluation codebook: strategies that respondents reported for judging whether the tool's output was acceptable, applied only when grounded in explicit textual evidence. $n$ (\%) is accounts in the analytic sample ($N = \NAnalytic$) carrying each code; quotes are verbatim answers to the evaluation prompt, reproduced as written.}
\label{tab:eval-codebook}\\
\toprule
Code & $n$ (\%) & Definition & Example account \\
\midrule
\endfirsthead
\multicolumn{4}{l}{\tablename~\thetable{} (continued)}\\
\toprule
Code & $n$ (\%) & Definition & Example account \\
\midrule
\endhead
\bottomrule
\endlastfoot

Run code & \NRunCode{} (\PctRunCode) &
Manually initiate code execution and check that it runs without producing
errors; includes phrases such as ``I tested it'' absent any indication of a test suite. &
``no errors, examining the output of the multiplication to confirm that it
was what I expected'' \\
\addlinespace[2pt]
Inspect output & \NInspectOutput{} (\PctInspectOutput) &
Run the code and examine the resulting outputs, such as interactively
inspecting a data object or looking at what is printed to console, file, or
logs; excludes inspecting plots or figures. &
``Its output data was similar to what I have been using.'' \\
\addlinespace[2pt]
Read code & \NReadCode{} (\PctReadCode) &
Read through the generated code, sometimes ``line by line''. &
``Reading through the output before using, and then trying the suggested
fix myself.'' \\
\addlinespace[2pt]
Inspect visualization & \NInspectViz{} (\PctInspectViz) &
Evaluate the output by visually examining a plot or other visual output
that the code produces. &
``I would run the chunk of codes in my local environment and check the
output, for example, checking whether the graph meet my requirement.'' \\
\addlinespace[2pt]
Domain knowledge / intuition & \NDomainKnowledge{} (\PctDomainKnowledge) &
Draw on the respondent's own intuition or domain knowledge. &
``reading it and seeing if I would have written same'' \\
\addlinespace[2pt]
Check reference & \NCheckReference{} (\PctCheckReference) &
Consult official documentation, a reference manual, published work, or a
reference code implementation. &
``Researched and used online documentation to confirm. I also wrote unit
tests'' \\
\addlinespace[2pt]
Compare to benchmark result & \NBenchmark{} (\PctBenchmark) &
Compare the AI-generated output against a previous result (a benchmark). &
``I ran the code and compared the outputs to known quantities.'' \\
\addlinespace[2pt]
Ask AI to explain & \NAskAI{} (\PctAskAI) &
Prompt the AI tool to explain the generated code or answer follow-up
questions about it. &
``I tried to read it thoroughly and also ask the tool to explain it to me
so I know that each step that it is doing is what I wanted to do.'' \\
\addlinespace[2pt]
Unit tests / test suite & \NUnitTests{} (\PctUnitTests) &
Check how code performs on a test suite (usually unit tests); applied only
when language specifically indicated a test harness. &
``Created a test bench and verified that the results generated by this
function is what I'm expecting.'' \\
\addlinespace[2pt]
Colleague review & \NColleagueReview{} (\PctColleagueReview) &
Check work with labmates, a professor, or another person with relevant expertise. &
``Double check with supervisor and colleagues'' \\
\addlinespace[2pt]
Check math by hand & \NMathByHand{} (\PctMathByHand) &
Manually calculate mathematical expressions to check that the code gives
the same result. &
``I test for edgecases, and do some easy calculations to check if it gets
the answer right.'' \\
\addlinespace[2pt]
Cross-check with another AI & \NCrossCheckAI{} (\PctCrossCheckAI) &
Use a different AI tool or model and compare outputs. &
``I run it and check the results. I also use other gen AI tolls to test
it'' \\
\end{longtable}

\endgroup

\begin{figure}[htbp]
  \centering
  \includegraphics[width=0.85\textwidth]{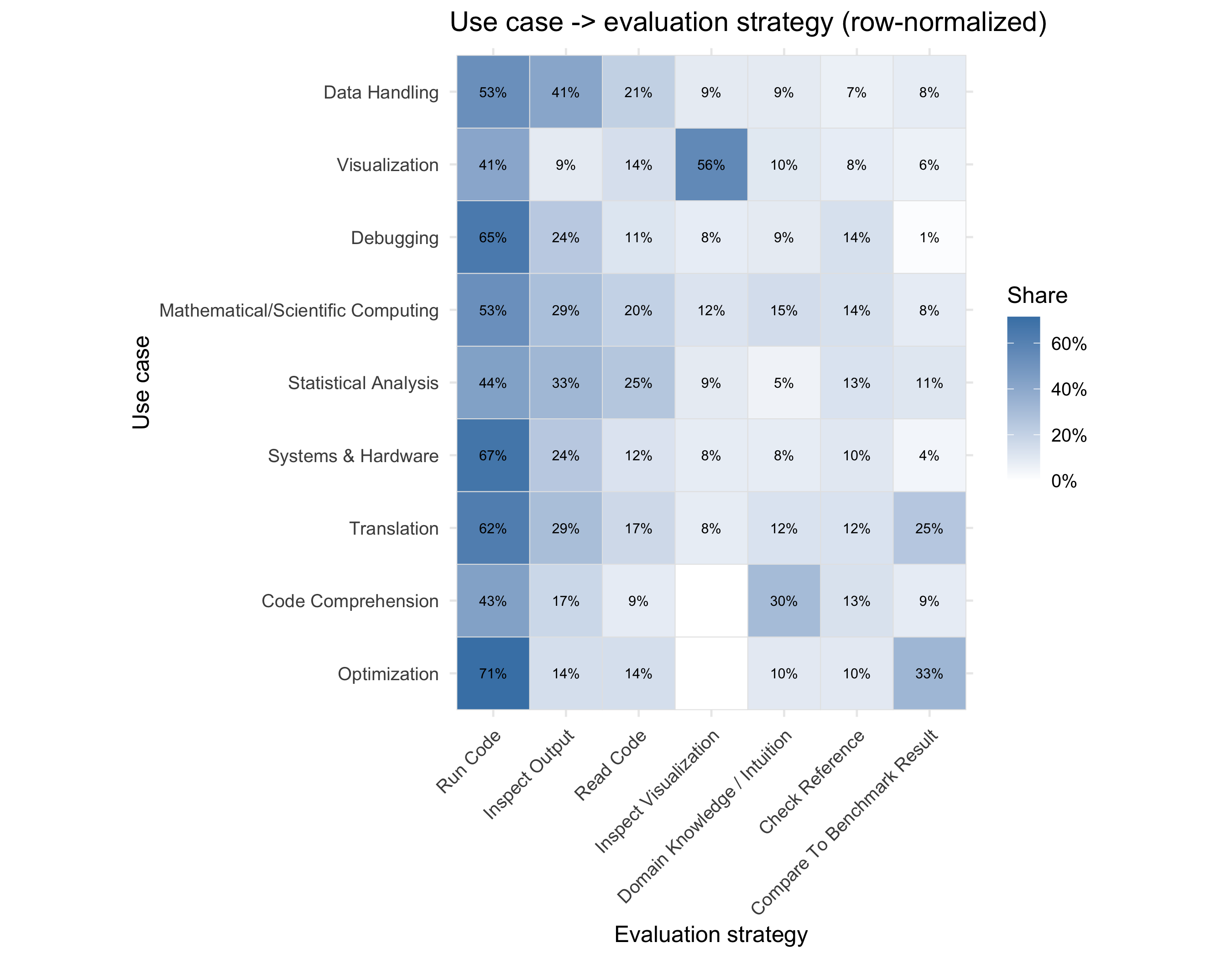}
  \Description{Heatmap of use cases (rows) by evaluation strategies (columns), each cell the share of accounts with that use case that reported that strategy.}
  \caption{Evaluation strategies conditioned on use case. Each cell is the
  share of accounts carrying the row's use-case code that also carried the column's evaluation code. Rows can sum to more than 100\% because accounts often reported several methods. Codes applied to
  fewer than 20 accounts are omitted, and rows and columns are ordered
  by overall frequency.}
  \label{fig:cooccurrence}
\end{figure}

We review some of the most common use cases, along with their most common evaluation strategies. The share of each use case's accounts reporting each evaluation strategy is summarized in Figure~\ref{fig:cooccurrence}.

\subsubsection*{Data handling}

Data handling, the most common use case, covered processing, cleaning,
deduplicating, reshaping, reformatting, merging, and harmonizing data, as well as converting between data types, usually in preparation for downstream analysis or reporting. A recurring task was loading data from the file system into an interactive Python, R, or MATLAB session, often
from specially formatted files produced by simulation software or by
measurement devices such as microscopes, telescopes, and biomedical imaging
instruments. Respondents also described complex filtering operations, in which they asked the tool to translate a set of logical filtering steps and conditions into implementations for common data handling libraries such as pandas. Within a programming environment, tasks
included converting between data structures (e.g., a table to an array, model output to a CSV file, strings to dates), merging tables, and applying
transformations or imputations.

Respondents' accounts suggest two main reasons they reached for generative
AI in this category. The first was the complexity of the data itself:
harmonizing data collected from multiple instruments with different sampling
rates, parsing specialized file formats tied to domain-specific
instrumentation, and contending with missingness and irregularity introduced
during collection. The second was unfamiliarity with the specific libraries involved. Respondents often knew what operation they wanted but not how to express it in the tool at hand, as when one respondent used ChatGPT to
work out how the Astropy package computes a coordinate transformation.
Sometimes this unfamiliarity reflected a working environment rather than a
knowledge gap: ``I could have written a SAS macro to perform this action,
but I'm the only person in my research group that uses SAS, meaning that if
I want any of my coding work to be reproducible by my team, or even usable
by my team, it has to be done in R. Everything I do in R takes a lot longer
due to having less experience in it, so I consulted [institutional AI tool\blinded{, name withheld for review}{}]''.

About half of data handling accounts described more than one validation
method, so a combination of strategies was typical. The most commonly
reported strategy was running the generated code (e.g., ``I tested the code
against actual data to be sure that it worked''), although only a minority of accounts (\NDhRunOnly/\NDataHandling) reported running code without any other strategy. Some respondents deliberately ran generated
code on samples or constructed inputs: ``I created two mock data files with
a small number of entries for which the answer was easily determined and ran
them through the code.'' It was not always clear from these accounts whether
execution served to confirm that the code ran without error or to check its
correctness against a known answer. Among respondents who explicitly
reported inspecting outputs, some named the qualities they examined --- ``if
it fully loaded the date ranges I was looking for''; ``I opened the dataset
and 1. first, checked if any value was imputed; 2. compared the imputed
values with other values within each column'' --- while others described
inspection only in general terms (``eyeballing the result,'' ``I got my
desired result''), leaving unstated which attributes of the output they
attended to. A smaller number of respondents described reading the code
itself, typically in combination with testing it: ``reading it over to make
sure there were no obvious errors, then testing it.''

\subsubsection*{Visualization}

Visualization was the second most common use case. Respondents used
generative AI both to produce less familiar chart types and to recall the
syntax of common plotting libraries. Much of the work was incremental refinement of plots that the respondent had already generated: adjusting colors,
working with heat maps, or layering an additional variable into an existing
graph. As one respondent put it, ChatGPT ``is especially useful when it
comes to completing menial programming tasks such as formatting a figure I
have already generated to include certain features that are beyond my
knowledge of default plotting.'' Others framed the tool less as a code
generator than as a faster route into documentation. One respondent
explained that ``much of the MATLAB (and Python) syntax for generating plots
is oddly specific. I would normally have to spend time digging through the
documentation to remember exactly what the call is to thicken the line plot
or find the hexadecimal color code\ldots{} With an AI tool, I can ask that
question and usually get an answer much faster than having to google it,''
adding, ``I'm not having ChatGPT write code so much as operate as a shortcut
to digest the documentation.'' The tasks ranged from publication-ready formatting (e.g., fonts, titles, labels, line thickness) to more substantive manipulation of how data was grouped or ordered for a plot. Respondents
differed in how much they specified up front: some named the library or plot
type they wanted, while others asked the tool to recommend one, as when a respondent used ChatGPT ``to recommend different plotting functions in both
libraries I was familiar with and libraries I had not heard of or
considered.''

The dominant evaluation strategy was inspecting the resulting graph,
reported in just over half of visualization accounts (\NVizInspectViz/\NVisualization). Where respondents specified what they looked at, they described checking the figure against their data and expectations: ``I double-checked the graphs' points with my data to see if
they make sense, and the trend was right based on my judgment''; ``the
results were slightly different than what I expected, so my next step was
looking into the violin plot API to understand what the keyword arguments
were doing.'' Several respondents treated visual output as almost self-validating. One wrote that ``it's entirely about the visualization so
it's literally visually validated,'' another that ``I visually determined it
was the chart I wanted,'' and a third that, for plot generation
specifically, ``I probably accept it immediately.'' This treatment of visual output as self-validating stands in some contrast to data handling, where the correctness of an output may not be apparent from the output itself.

\subsubsection*{Debugging}

Debugging was the third most common use case, and the accounts here often centered on working with unfamiliar tools. Respondents frequently reached
for generative AI when confronting a language, framework, or version they
did not know well: ``sometimes there are errors in C++, for example, that I
am totally unfamiliar with because I come from a primarily Python-based
background, and AI helps me narrow down the cause''; ``I recently started
coding models in JAGS in R, and it's pretty common to run into compilation
errors due to syntax mistakes, or improperly specified priors.'' Version
migration was a recurring trigger, as in a respondent adapting a script
across MATLAB releases: ``This code worked in MATLAB 2022 but fails in
MATLAB 2025. Can you help me find which command or function is causing the
error?'' The prototypical workflow was to paste an error message and ask the tool to explain or resolve it. The errors often arose from data handling, analysis, or visualization work, and sometimes from systems-level issues such as CUDA setup or portability across operating systems. Some respondents noted that they turned to AI only after exhausting their usual resources. One described Stack Overflow as ``always the first place I look for help'' and consulted ChatGPT only when that failed. Beyond syntactic
errors, a few respondents used the tool for more semantic debugging,
including one account of locating a substantive error in a paper under review: ``I accessed their code base, and it was about 2000 lines. It
would be hopeless to locate the error. I described what was wrong in the
paper, put their code into ChatGPT and it \emph{instantly} located the exact
error.''

The dominant evaluation strategy, reported in most debugging accounts (\NDbgRunCode/\NDebugging), was running the code, typically to check whether the original error had been resolved. Respondents described
accepting a fix when the code ``did not throw an error and still did what I
wanted it to do,'' or, more simply, when ``I run it in R and see if I get
the results I want.'' Debugging stood out as the use case where running code
most often appeared alone: \NDbgRunOnly{} of \NDebugging{} accounts
reported it as the sole evaluation strategy, and only about a third
(\NDbgMulti/\NDebugging) described more than one strategy overall. Some
respondents, however, went beyond confirming that the error had cleared.
They read the suggested code line by line and cross-referenced unfamiliar recommendations against documentation or forums: ``if the error is caused by something I'm less familiar with, I'll cross-reference what UM-GPT suggests with a JAGS forum or Stack Overflow posts.'' Several described a personal rule against pasting generated code directly, instead retyping or adapting it themselves.

\subsubsection*{Mathematical/scientific computing}

Mathematical/scientific computing covered use cases in which the core
task was implementing or solving a mathematical or domain-specific
scientific problem: finding solutions to systems of equations, translating a
set of governing equations into working code, writing simulations, and
setting up numerical optimization. Some respondents described the task in terms of the mathematics itself, as in ``find solution for a nonlinear equation system'' and ``solving equations,'' while others were implementing a
specific model from their domain, such as a drift-diffusion model for plasma
discharge or a simulation in item response theory. A recurring pattern was
starting from a known formalism, typically a set of equations drawn from a
published article or textbook, and asking the tool to render it as code. One
respondent working on an underdetermined nonlinear ordinary differential equation (ODE) noted that Claude
``provided several feasible approaches to solve my problem'' and ``vastly simplified the algebra needed to arrive at a final usable expression in minutes --- something which would have taken me a week at least.'' In this category, then, the tool was sometimes used for the mathematics as much as for the code.

Respondents in this category especially often combined strategies. Most accounts described more than one strategy (\NMscMulti/\NMathSci), and only \NMscRunOnly{} relied on running the code alone. Running the code was
still the most common single strategy (\NMscRunCode/\NMathSci), but respondents also used other strategies. Some compared the output
against an independent standard: published results and textbook examples or
``output products of previous versions of the pipeline'', a kind of
computational benchmark. Others verified the underlying mathematics directly rather than trusting the implementation, by checking a derivation step by step (``I asked it to show me steps that I verified''), by confirming that a hand-computed quantity matched the code's output, or, in one case, by cross-checking ``a few hand-computed current balances'' against the tool's
outputs ``to confirm physical feasibility.'' Some respondents chose strategies to avoid engaging in mathematical justification directly: one described implementing ``a new function with heavy math that I didn't fully
understand, and didn't need to fully understand,'' and validated it entirely
behaviorally, by creating a ``test bench'' and confirming that the function
returned the results they expected.  Several respondents also drew on domain knowledge or expert judgment as the final check, assessing
whether ``the results/scientific figures are scientifically sound'' or
checking the output with a supervisor, colleagues, or PI.

\subsubsection*{Statistical analysis}

Statistical analysis captured use cases in which the goal was to apply a statistical or machine learning method to find patterns in data: fitting regressions, running statistical tests, calculating descriptive statistics, and implementing machine learning pipelines. In many cases, respondents had clear analysis strategies in mind but needed to learn less familiar libraries or functions. For example, one respondent shared a direct prompt that they had written: ``I have the data and script written in SPSS for the
mediation models. Can you help me quickly import data into R and format as
needed to do proposed analysis using lavaan.'' Several described doing
their first-ever analyses with \texttt{pandas} or a network analysis
package.

Other respondents knew the analysis that they wanted but not whether an implementation existed, and used the tool to survey available packages:
``Are there R packages to implement techniques like King and Zeng's
methods?'' Another ``asked it what built in change-point functions existed
in R, what their use-cases were, and what arguments each function used.''

Others needed to customize the structure of statistical models in clearly
specified ways, like converting confidence intervals from analytical to
bootstrap, re-estimating a returns model at daily rather than monthly
frequency, or matching model structure to the specifics of their dataset.

Some responses showed scientists using AI tools as recommender systems for analysis strategies. One respondent prompted, ``What kind of
statistical analysis can I do in R that can help me find significant
insights into the connection between [variables in dataset]?'' Another
reported, ``I described the data patterns to the tool and asked what
methods were available for detecting such patterns by
clustering\ldots{} It suggested a package for k-means clustering and
thought that implementing with dynamic time-warping would be up to the
task\ldots{} I wanted to know how to identify the optimal number of
clusters, and it suggested elbow plots.''

Evaluation strategies were varied and frequently combined
(\NAnaMulti{} of \NAnalysisUC{} accounts described more than one). Running the code
was the most common strategy (\NAnaRunCode/\NAnalysisUC), though only \NAnaRunOnly{} reported
running code as their sole check (``If it works without bugs or not'';
``try to run and check any error is coming or not''). One respondent made
their multi-stage strategy explicit: ``I tested the output in R to first
see if it ran and then I looked through the output for indications of
successfully completing the intended task (correct values, missing values,
etc.).'' When inspecting output (\NAnaInspectOutput/\NAnalysisUC), respondents described
checking for specific failure modes and properties: variables ``not
transformed to the correct scale,'' asymmetry in a matrix that should be
symmetric, or implausibility against an internal sense of what a reasonable
result looks like (``I pulled out numbers and saw if the classification
looks ok''; ``I check if the results look reasonable and match what I
expect''). The most involved evaluations piloted generated code on data with known answers or re-derived the result independently. One respondent used a ``much smaller dataset with known regression result'' as a reference, and another ``also ran one model without the function and compared the results to ensure they were the same.''
\subsubsection*{Evaluating AI output}

Across the corpus, the strategies that respondents reported were dominated by the lightweight checks noted above (Figure~\ref{fig:code-counts}, right): running the generated code and inspecting its output, followed by reading the code, inspecting a visualization, and drawing on domain knowledge or intuition. Among these common strategies, respondents reported a wide range of diligence. Some were emphatic
about scrutiny (``I never mindlessly copy-paste''), while others acknowledged that scrutiny was the first practice they dropped under time pressure: One respondent admitted that ``the longer I'm working on a code/the more
complicated it gets, the less I read it and the more I resort to just
running the code and seeing what happens.'' Another described reviewing
code only after using it: ``This usually comes after checking if the
code works, however, so I guess I am frequently running untested code.''

The remaining
strategies were comparatively rare. Respondents consulted external
references such as documentation or forums in \NCheckReference{}
accounts (\PctCheckReference\%) and, in \NBenchmark{} (\PctBenchmark\%), compared output against an independent benchmark such as published results, a prior pipeline, or a known-answer dataset. More formal or
collaborative checks were rarer still: some form of automated testing
appeared in only \NUnitTests{} accounts (\PctUnitTests\%), asking the
tool to explain its own output in \NAskAI{} (\PctAskAI\%), review by another person in \NColleagueReview{} (\PctColleagueReview\%),
checking the mathematics by hand in \NMathByHand{} (\PctMathByHand\%), and
cross-checking against a second AI tool in \NCrossCheckAI{}
(\PctCrossCheckAI\%). When review was described, the language suggests an ad-hoc assembly of nearby reviewers rather than a dedicated code review process (``usually double check with professors''; ``I shared the code with other people and asked them to test it as well''). 

Overall, validation was seldom externalized or formalized. In most accounts, the person who prompted
for the code also ran it, inspected the result, and decided whether it was acceptable. Independent reviewers and automated test suites each appeared in only a small fraction of cases. These counts reflect the evaluation strategies that respondents considered
worth reporting, not necessarily everything they did
(Section~\ref{sec:limitations}).

\subsection{Quantitative analysis}
\label{sec:quant}

\subsubsection*{Who uses AI for what?}

We first asked whether reported use cases and evaluation strategies varied
with respondents' years of programming experience and research area.

\paragraph{Experience.} For each of the top five use-case codes, we compared
log-transformed programming experience between respondents whose
account did and did not receive that code (Welch's t-tests,
$N = \NExpUseCaseTests$). The largest difference was for debugging: Respondents describing debugging use cases were somewhat less experienced
($M = \MeanLogDebug$ vs.\ $\MeanLogNoDebug$ log units, roughly \YearsDebug{}
vs.\ \YearsNoDebug{} years), but this did not reach significance
($t(\DfDebugExp) = \TDebugExp$, $p = \PDebugExp$, BH $p = \PBhDebugExp$).

We also tested whether the top five evaluation strategies were related to programming experience, but saw little evidence of a relationship. In per-strategy comparisons ($N = \NExpStrategyTests$), respondents who reported a given strategy did not differ in programming experience from those who did not. The largest difference was for reading code
($t(\DfReadExp) = \TReadExp$, $p = \PReadExp$, BH $p = \PBhReadExp$). The
number of distinct strategies a respondent reported (counted over all twelve
evaluation codes) was likewise unrelated to experience
($r(\DfStrategyCountExp) = \RStrategyCountExp$, $p = \PStrategyCountExp$).

\paragraph{Research area.} Two use cases differed in frequency by research
area in per-code chi-square tests (5 areas $\times$ code present/absent).
Mathematical/scientific computing use cases were most common among engineers
(\PctMscEngineering\% of engineering accounts vs.\ \PctMscLifeSci\% in the life sciences, \PctMscPhysSci\% in the physical sciences, and \PctMscOther--\PctMscSocialSci\% in other areas and the social sciences;
$\chi^2(\ChisqMscDf) = \ChisqMsc$, $p = \PMscArea$, BH $p = \PBhMscArea$).
Statistical analysis use cases showed roughly the reverse pattern (\PctAnaOther\% of
Other and \PctAnaLifeSci\% of Life sciences accounts vs.\
\PctAnaEngineering--\PctAnaPhysSci\% in Engineering and Physical sciences;
$\chi^2(\ChisqAnaDf) = \ChisqAna$, $p = \PAnaArea$, BH $p = \PBhAnaArea$).

In sum, neither reported use cases nor evaluation strategies varied
substantially with programming experience. The differences that we could detect were associated with research area, and in expected directions: mathematical
and scientific computing was more common among engineers, while statistical
analysis use cases concentrated in the life and social sciences.

\subsubsection*{How is experience related to confidence?}

\begin{figure}[htbp]
  \centering
  \includegraphics[width=\textwidth]{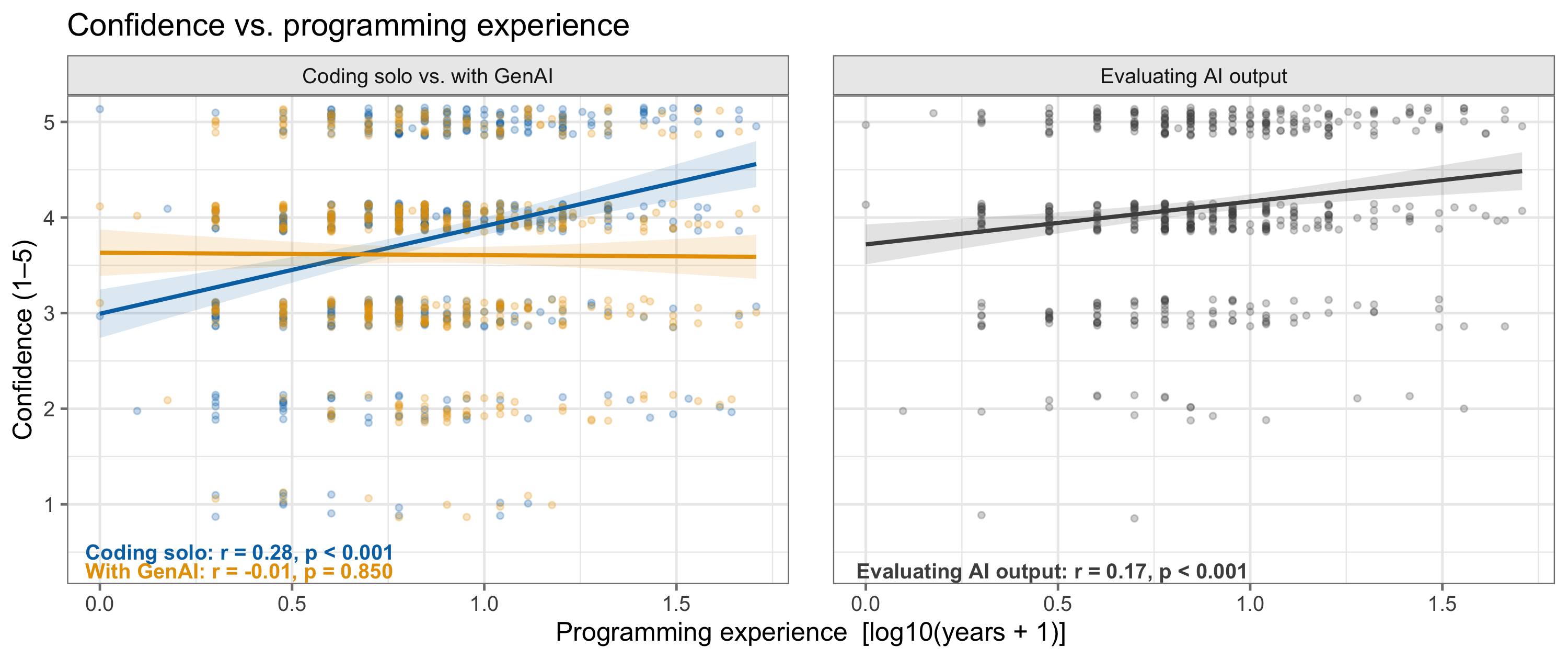}
  \Description{Two scatter panels of confidence ratings against log programming experience with fitted lines: solo and GenAI confidence on the left, evaluation confidence on the right.}
  \caption{Confidence ratings (1--5) as a function of programming
  experience ($\log_{10}(\text{years}+1)$). Left: confidence completing the
  described task without generative AI (``coding solo'') and confidence in
  the GenAI tool. Right: confidence evaluating the tool's output. Points are individual accounts (vertically jittered). Lines are linear fits with 95\% confidence bands.}
  \label{fig:confidence-experience}
\end{figure}

Respondents rated their confidence in (a) completing the task independently
(``solo''), (b) completing it with the GenAI tool (``GenAI''), and (c)
evaluating whether the tool's output was correct (``evaluation'').
Figure~\ref{fig:confidence-experience} plots each rating against
programming experience. For the confidence analyses we included every account with confidence ratings, even those that could not be assigned qualitative codes. Solo confidence rose with programming experience ($r = \RSoloExp$, $p = \PSoloExp$, $N = \NSoloExp$), as did evaluation confidence
($r = \REvalExp$, $p = \PEvalExp$, $N = \NEvalExp$). GenAI confidence, by
contrast, was unrelated to experience ($r = \RGenaiExp$, $p = \PGenaiExp$,
$N = \NGenaiExp$). This flat relationship could reflect that we asked participants to describe a recent use case. Many will have no use case to report in which they were very unconfident in the AI, because they would not have used it at all.

This divergence is easiest to see in the within-respondent gap between solo
and GenAI confidence (solo $-$ GenAI). The gap widened with experience ($r = \RGapExp$, $p = \PGapExp$, $N = \NGapExp$), and the fitted line crossed zero at roughly four years of programming experience
($\log_{10}(\text{years} + 1) = \GapCrossLog$,
$\approx \GapCrossYears$ years). In other words, less experienced programmers tend to use AI for tasks that they are not as confident they could do alone, while more experienced programmers tend to use it for tasks that they are confident they could do themselves. 

\subsubsection*{What use cases are respondents most confident about?}

We next asked whether the use case itself was related to any of the three
confidence ratings, over and above programming experience (a natural
confound, given the associations above). For each confidence rating, we first compared a linear model with experience alone to a model adding all five use-case indicators as a block (nested F-test). This omnibus test asks
whether the use cases jointly explain additional variance.

Adding use cases improved the model for evaluation confidence
($F(\FUcEvalDfNum, \FUcEvalDfDen) = \FUcEval$, $p = \PUcEval$,
$\Delta R^2 = \DeltaRsqUcEval$) but not for solo confidence
($F(\FUcSoloDfNum, \FUcSoloDfDen) = \FUcSolo$, $p = \PUcSolo$) or GenAI
confidence ($F(\FUcGenaiDfNum, \FUcGenaiDfDen) = \FUcGenai$,
$p = \PUcGenai$).

\begin{figure}[htbp]
  \centering
  \includegraphics[width=\textwidth]{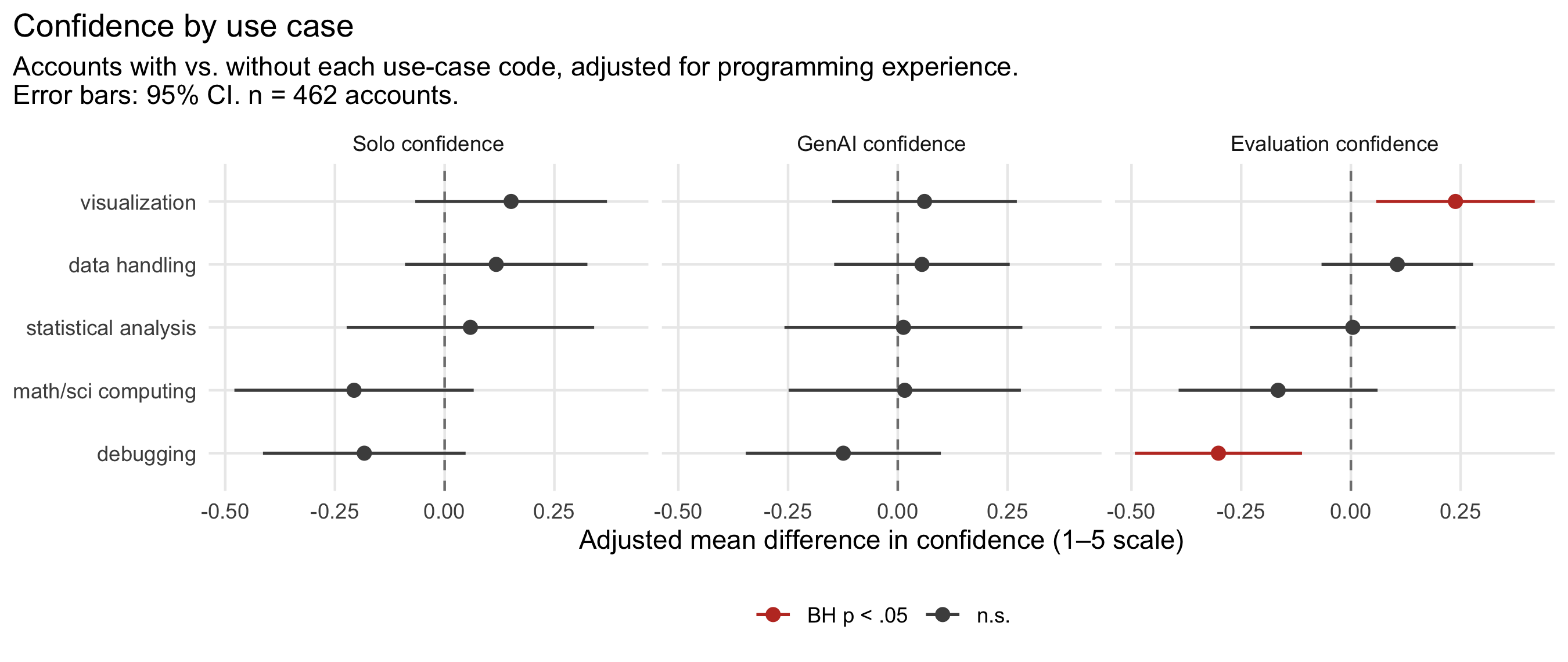}
  \Description{Forest plot of experience-adjusted mean differences in three confidence ratings for each of the five main use cases, with 95 percent confidence intervals.}
  \caption{Confidence by use case. Points are experience-adjusted mean
  differences in each confidence rating (1--5 scale) between responses with
  and without each of the five most common use-case codes, from linear
  models of confidence on log-experience plus the use-case indicator (one
  model per code and rating). Error bars: 95\% CIs. Red marks differences with BH-adjusted $p < 0.05$ within each confidence rating. $n = \NAncovaFig$
  responses with complete confidence and experience data.}
  \label{fig:confidence-usecase}
\end{figure}

For evaluation confidence, we then fit per-code models adjusting for
experience (each an ordinary linear model of confidence on experience plus one use-case indicator). The use-case coefficient is the experience-adjusted mean difference between accounts with and without that code.
Evaluation confidence was lower for debugging use cases
($M = \MeanEvalDebug$ vs.\ $\MeanEvalNoDebug$; adjusted difference
$b = \BEvalDebug$, $t(\DfAncova) = \TEvalDebug$, $p = \PEvalDebugAnc$, BH
$p = \PBhEvalDebugAnc$) and higher for visualization use cases
($M = \MeanEvalViz$ vs.\ $\MeanEvalNoViz$; $b = \BEvalViz$,
$t(\DfAncova) = \TEvalViz$, $p = \PEvalVizAnc$, BH $p = \PBhEvalVizAnc$;
Figure~\ref{fig:confidence-usecase}).

In sum, the use case that a respondent described was related only to their evaluation confidence: Respondents were more confident evaluating AI output for visualization and less confident for debugging. Neither
confidence in the tool nor confidence in completing the task alone varied
with the use case reported.

\subsubsection*{How does evaluation strategy relate to confidence?}

Finally, we asked whether evaluation confidence tracks respondents' reported
evaluation strategies. Adding the five most common evaluation codes as a block
did not improve on an experience-only model of evaluation confidence (nested
F-test: $F(\FVmEvalDfNum, \FVmEvalDfDen) = \FVmEval$, $p = \PVmEval$). In
univariate t-tests, respondents who reported inspecting visualizations
($M = \MeanEvalWithViz$ vs.\ $\MeanEvalNoStratViz$;
$t(\DfEvalStratViz) = \TEvalStratViz$, $p = \PEvalStratViz$) or reading code
($M = \MeanEvalWithRead$ vs.\ $\MeanEvalNoStratRead$;
$t(\DfEvalStratRead) = \TEvalStratRead$, $p = \PEvalStratRead$) reported higher evaluation confidence, but neither effect survived correction for multiple comparisons (both BH $p = \PBhEvalStratViz$). The number of distinct strategies that a respondent reported was also unrelated to evaluation confidence ($r(\DfStrategyCountEval) = \RStrategyCountEval$, $p = \PStrategyCountEval$). In a model predicting strategy count from evaluation confidence with experience as a covariate, $b = \BStrategyCountEval$ strategies per confidence point ($p = \PStrategyCountEvalLm$).

Among the measures we collected, the strongest correlates of evaluation confidence were confidence in completing the task alone ($r = \REvalSolo$,
$p = \PEvalSolo$) and confidence in the GenAI tool to complete the task
($r = \REvalGenai$, $p = \PEvalGenai$). These two ratings were uncorrelated with each other ($r = \RSoloGenai$, $p = \PSoloGenai$), so this pattern is not consistent with a single, underlying general confidence
factor. Both relationships are much stronger than any that we observed with reported evaluation strategies.

\section{Discussion}
\label{sec:discussion}

\subsection*{Summary of findings}

We coded accounts of generative AI use in research programming for two kinds
of information: the task that motivated the request, and the strategies used
to evaluate what came back. Use was concentrated in five tasks: data
handling, visualization, debugging, mathematical/scientific computing, and
statistical analysis, which together appeared in about three-quarters of
accounts where a use-case code could be applied. Evaluation across all of them was
dominated by informal, individual
checks. Over half of respondents reported running the generated code as an
evaluation strategy. 

Quantitatively, neither use cases nor evaluation strategies varied much with programming experience. The few differences we could detect were by field,
in expected directions. Confidence, by contrast, did vary with experience. Less experienced programmers reported more confidence
in the AI than in themselves, and more experienced programmers the reverse,
with the crossover at roughly 4 years of experience. One interpretation is that novices may use these tools to attempt work that they could not do alone, while experienced programmers use them to save time on work that they could do themselves. Finally, we did not detect a relationship between respondents' confidence in their evaluations and either the evaluation strategies that they reported or the number of strategies they reported. In our data, evaluation confidence was associated with programming experience, use case (higher for visualization, lower for
debugging), reported confidence in doing the task alone, and reported confidence in the tool to complete the task. 

\subsection*{Interpretation}

These results describe how scientists evaluate AI-generated code, but they
do not license a strong normative claim that people are under- or over-evaluating, because the appropriate level of evaluation is not fixed across
tasks or contexts. Instead, scientists appear to make situated judgments about what constitutes sufficient evidence of correctness based on the task, the observability of its output, and their own knowledge of the underlying data, code, and scientific domain. For a cosmetic change to a plot, visual inspection may be exactly the appropriate check. The respondents who described figures as ``literally visually validated'' may be correctly, not carelessly, calibrated. For data handling, correctness is often not visible in the output itself, and ``it ran'' establishes little. Debugging is the hardest to judge from the outside: Are respondents making small syntactic repairs
that never touch the scientific assumptions in the code, or doing things
until an error message goes away, possibly altering parts of the code they
should not in the process? Our accounts cannot distinguish these readings, and the two
have very different implications.

The finding that evaluation confidence was not significantly related to reported evaluation strategy has two candidate explanations. One is methodological: Free-text accounts are a coarse instrument, and respondents may under-report checks that they actually did. For example, someone who writes ``I ran it'' may also have compared the output against expectations without saying so. We may have limited statistical power to detect a real relationship. The other is that
confidence in evaluating AI output is not primarily a product of evaluation strategies for many respondents. That evaluation confidence was associated with confidence in the tool is consistent with trust partially substituting for verification, consistent with the finding of \citet{Lee2025TheWorkers} that knowledge workers with higher
confidence in generative AI engaged in less critical evaluation. Our data
cannot adjudicate between these readings, and both may be partly true.

The association between confidence and trust does not imply that highly confident respondents evaluated poorly, nor can our data establish the quality of any individual evaluation. It does suggest that subjective confidence should not be treated as a proxy for evaluation strategies.

In nearly every account, the respondent described a closed loop of
interaction between themself and their AI tool. Review by another person appeared in \PctColleagueReviewWhole\% of accounts and automated
testing in \PctUnitTestsWhole\% (though this does not mean only \PctUnitTestsWhole\% of scientists use testing, only that it was not mentioned explicitly in most accounts). This closed loop is consistent with earlier surveys reporting low use of peer code review and testing among scientists \citep{Carver2022AStates, Hannay2009HowSoftware}. We would not expect
this infrastructure to appear quickly now that generative AI tools are available. Additionally, many of the use cases described could be challenging to fit into a typical test suite, or may not have sufficient epistemological weight to justify the overhead of testing in a formal sense (as tests accumulate technical debt, too). For example, some kinds of visuals and descriptive tables are rendered quickly to support exploration and hypothesis formation \citep{SutherlandKeller2025ResearchSoftware, Kery2017ExploringProgramming}. Errors here still matter (for example, filtering data in an unexpected way before summarizing or visualizing it), but repeated interaction offers many opportunities to notice such a problem. Qualitatively, we observe wide variation, ranging from quick checks that no errors arise, to comparing results against internal expectations of reasonable values, to checking that other code implementations produce the same result, to writing test cases. This variation places the weight on individual scientists' expertise for judging both (a) what level of scrutiny generated code deserves in the research context and (b) whether that threshold is met in a given use case.

Tasks also differ in whether independent evidence is available. Scientists may be able to compare generated results against a reference such as a known quantity or a previous implementation, while in other cases no convenient reference exists. A more useful distinction than formal versus informal evaluation is whether the evidence stays inside the immediate human--AI interaction or introduces an independent point of comparison. Running generated code, visually inspecting its output, and asking the same AI to explain its response are all useful, but they preserve the same interaction loop. Checks such as a reference implementation, a hand calculation, or review by another person introduce evidence that is at least partially independent of the generation process. This distinction may offer a better basis for supporting evaluation than encouraging greater use of conventional software testing.

\subsection*{Implications}
If the burden of devising verification activities during a session, and then executing them, falls mostly to individual scientists, we see several ways to support them. Tools might default to offering to produce lightweight test datasets for interactively running through any generated code. For example, to verify an AI-generated data handling step, tools might offer a small, easy-to-inspect sample dataset to run through the code so that the output can be observed directly (a strategy several respondents already reported using, just manually). With agentic workflows growing in popularity, tools could also \textit{report to the scientist} that an agent is running this test case and then interactively show the result for further iteration.

Beyond data handling, AI programming tools could generate verification artifacts alongside code, tailored to the task. For visualization, this means diagnostic views that expose transformations applied before plotting. For debugging, a regression case shows that a fix addresses the original failure without changing unrelated behavior. For mathematical or scientific computing, the artifact is a comparison against a known quantity, a limiting case, or an alternative implementation. The goal is to reduce the burden on scientists of devising verification activities themselves, without prescribing a single notion of correctness.

Training scientists in the future could focus on modeling the informal evaluation strategies used by more senior scientists: After running generated code, what specific qualities of the output does a more experienced scientist inspect? What diagnostic plots do they make to understand the behavior of code that they did not write? Our interpretation of this data points toward scientists acting as calibrated instruments, that is, making small but important internal comparisons between what they expect of a result, given their knowledge of the data and the domain, and what actually happens. Such training may involve calibrating scientists' internal ``priors'' for how likely a given computational result is, and teaching them what steps they would take to re-calibrate when moving to an unfamiliar problem. 

Interfaces could also make the basis for acceptance more explicit. Rather than relying on a user's sense that an output looks correct, an AI-assisted programming environment might show what has been checked, such as whether the code ran, whether outputs were compared against known values, or whether an independent reference was consulted. Making those checks visible could help separate confidence from evidence without requiring every research programming task to adopt heavyweight software engineering practices.

\subsection*{Limitations}
\label{sec:limitations}

Our data consists of self-reported accounts of single recalled episodes, and by design
we conducted a content analysis bounded by what respondents wrote. As noted in the Interpretation section, respondents may have under-reported checks they performed, so our counts of
evaluation strategies are lower bounds on what was done. We
cannot know what people did, only what they remembered and chose to report. The survey is a
convenience sample, concentrated in U.S. higher education and likely to
overrepresent scientists interested in programming enough to answer a survey
about it. Confidence items were tied to a specific, recalled episode, and
the flat relationship between GenAI confidence and experience may partly reflect selection, because people rarely have use cases to report for tools that they do not trust.

Three further limitations concern how the accounts were produced and coded.
First, the survey asked scientists to describe their own diligence, so social
desirability may have led some respondents to over-report checks, and others,
writing briefly, to under-report them. Second, each account describes one
episode, so person-level claims (for example, that a respondent who described
a visualization task did not debug) are not licensed by the data, and our
quantitative comparisons treat one recalled episode as representative of the
respondent. Third, we designed the survey and coded the responses
ourselves. We mitigated the risk of confirming our own expectations through
two rounds of independent coding by all three authors, reliability checks
(Krippendorff's $\alpha = \KAlphaPairwise$ pairwise), and consensus
resolution of flagged accounts, but most of the corpus was coded by a single
author after the codebooks were finalized, and the coders were not blind to
the study's aims. Because the instrument was in English and the sample is U.S.-based, the practices described here may not transfer to
other research settings.

\subsection*{Future research}

These findings describe a moment in computing that is already past. Our data was collected in 2025, and most respondents were not using
agentic tools that can execute code, repair errors, and write their own test
suites. As tools change, we expect many of the use cases that we identified to persist, because they arise at points of friction between scientists and their tools, such as opaque error messages, switching between libraries with various degrees of familiarity, or requiring highly custom code for the particulars of data. But
evaluation may look quite different in an agentic workflow: Agents can design and execute certain kinds of validation work themselves, while also
producing far more code outside the scientist's direct oversight. If agents attempt validation work on their own, scientists may shift from designing their own verification strategies to judging whether the agent's strategies are appropriate. Where test suites are appropriate, agents might also help scientists write and automate them, although whether this convention takes hold is an open question. Studying validation work in the coming years will likely require observing scientists as they work with their tools, because their checks may not be legible in code or chat logs. For these reasons, we expect the scientific correctness of AI-assisted
research code to continue to rest on the scientist's skill in building a
well-calibrated mental model of the data they study, the computations they
apply, and the results they should expect.

\section*{Data Availability}

The survey responses cannot be shared, because the terms of the informed
consent under which they were collected do not permit redistribution, and the
written accounts contain free text that could identify respondents. Scripts
for conducting the quantitative analyses and generating the figures will be
provided upon publication. The full codebooks, with definitions
and example excerpts for every code, are given in Tables~\ref{tab:usecase-codebook} and~\ref{tab:eval-codebook}.

\begin{acks}
The lead author (GO) is supported by a grant from the Alfred P. Sloan Foundation.
\end{acks}

\bibliographystyle{ACM-Reference-Format}
\bibliography{references}

\end{document}